\documentclass[slac_one]{revtex4}
\usepackage{graphicx}
\usepackage{fancyhdr}
\begin{document}

\title{{\small{Hadron Collider Physics Symposium (HCP2008),
Galena, Illinois, USA}}\\ 
\vspace{12pt}
BeamLine Design for MINERvA TestBeam Detector} 

\author{C. E. P\'erez} \affiliation{Pontificia Universidad Catolica del Peru, Lima, PERU}
\author{A. M. Gago}    \affiliation{Pontificia Universidad Catolica del Peru, Lima, PERU}
\author{J. Morfin}  \affiliation{Fermi National Accelerator Laboratory, Batavia, IL 60510, USA}
\author{D. A. Jensen}  \affiliation{Fermi National Accelerator Laboratory, Batavia, IL 60510, USA}
\author{R. Gran}    \affiliation{University of Minnesota Duluth, Duluth, MN 55812, USA}

\begin{abstract}
The MINER$\nu$A TestBeam Detector calibrations will take place in the MTEST facility at Fermilab. It will use a beam of hadrons between 300 and 1500 MeV/c to analyze the response of the MINERvA detector components. The aim of the present work is to design the beam while considering radiation hazards at the hall. To accomplish this, we are using a Monte Carlo simulation based on Geant4 and actual data taken from measurements of the elements that make up the beamline.
\end{abstract}

\maketitle
\thispagestyle{fancy}

\section{The MINER$\nu$A Experiment}
MINER$\nu$A stands for {\bf M}ain {\bf IN}jector {\bf E}xpe{\bf R}iment $\nu$ - {\bf A} and is a high-statistics neutrino scattering experiment that will run in the NuMI Beam Hall at Fermilab. The experiment will use an intense and well-understood $\nu$ beam and an active fine-grained detector to collect a large sample of $\nu$ and $\overline{\nu}$ scattering events. The aim is to study a wide variety of processes such as precision measurements of the quasi-elastic $\nu$-A cross section, examination of nuclear effects in neutrino interactions, characterization of the transition from resonance to Deep Inelastic Scattering, measurements of the parton distribution functions at high $x_{BJ}$, etc. A more exhaustive reference to the physics of MINER$\nu$A can be found in \cite{minervaref}.

\section{The TestBeam Detector}
\begin{figure*}[t]
\centering
  \begin{tabular}{c c}
  \includegraphics[width=0.5\textwidth]{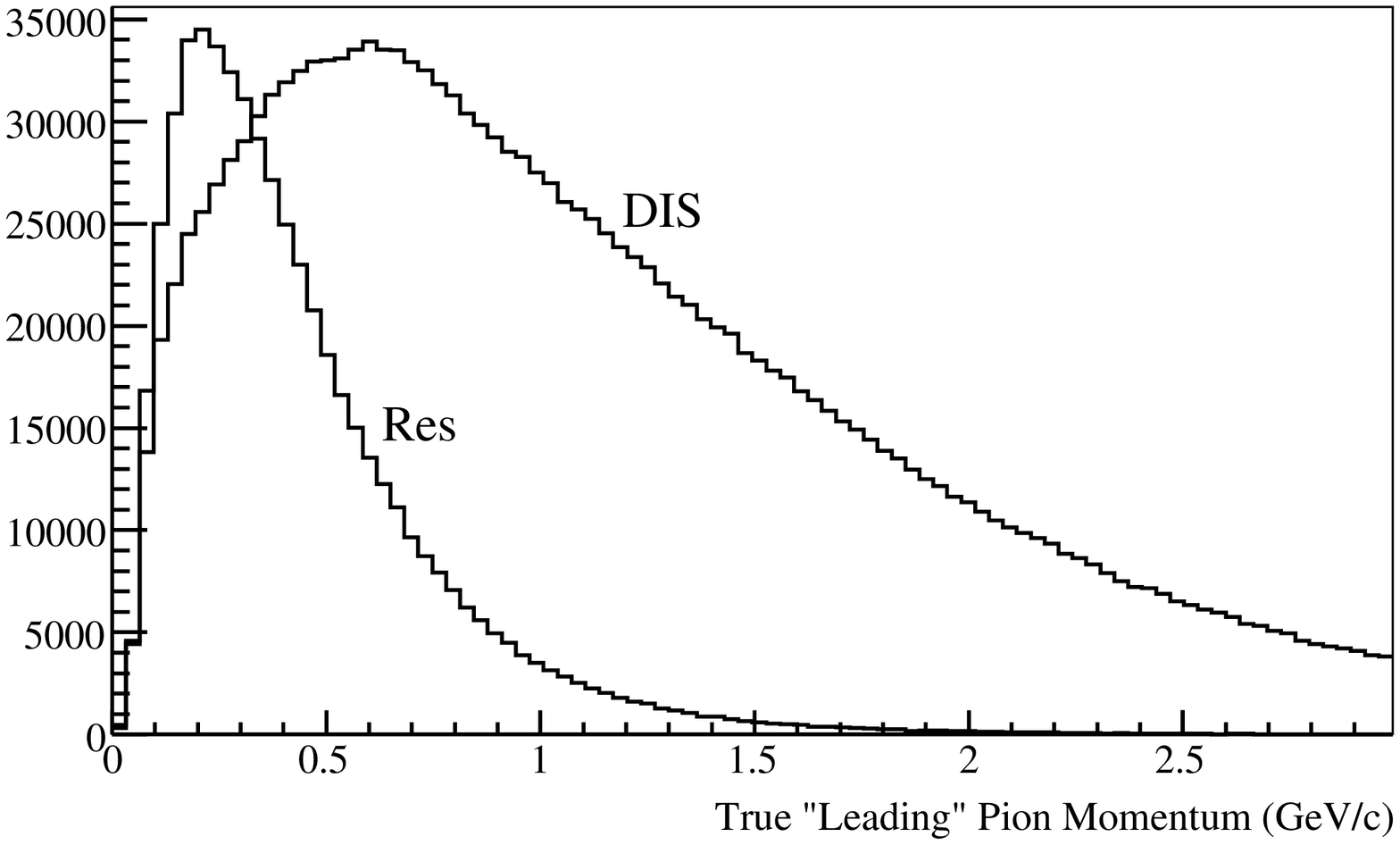}
  &
  \includegraphics[width=0.5\textwidth]{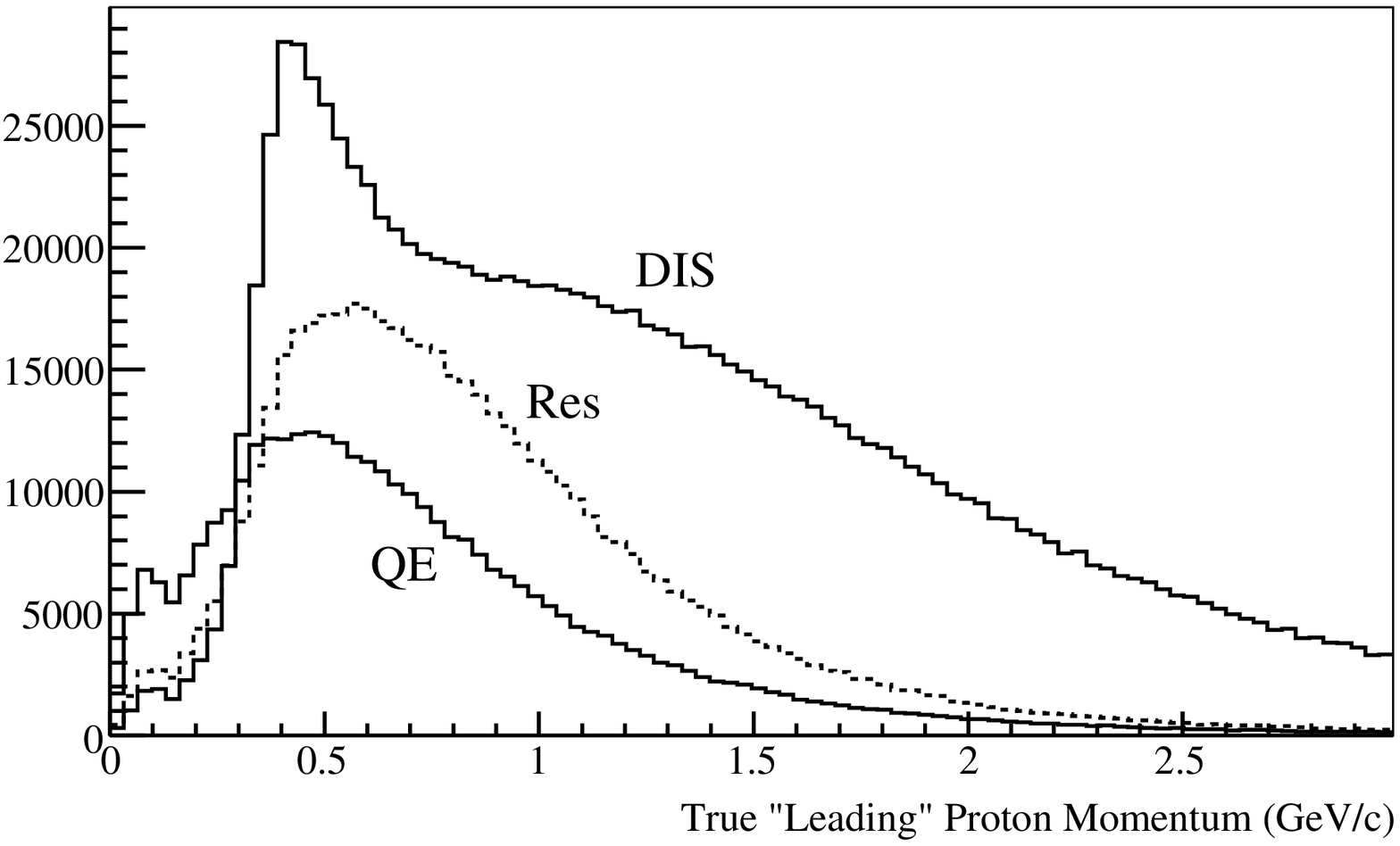}
  \end{tabular}
  \caption{Expected distribution of the leading pion (left) and proton (right) for quasi-elastic, resonance, and deep inelastic scattering events.  These estimates for spectra and number of hadrons are for a scintillator target and a medium energy configuration of the NuMI beam with $12\,10^{20}$ protons on target.} \label{figRik}
\end{figure*}

To understand the dynamics of the neutrino interaction, MINER$\nu$A has to be able to reconstruct the kinematics of the products with sufficient accuracy. Thus, the track of hadrons within the active detector has to be well determined. \\ The {\bf TestBeam Detector} (TBD) is a reproduction of the full detector in small scale. Both detectors will share the same resolution, but, due to its size, the TBD will allow various configurations of the calorimeter plates so that we can reproduce different sectors of the main detector. We will have to test at least three modes: electromagnetic, hadronic and tracker. One of the most important purposes is to provide a calibration of the calorimetric response to hadrons that pass through the test beam detector's fiducial volume. To this end the TBD will be exposed to a hadron beam in the energy range we expect for the products of the neutrino interactions with nuclei; see figure \ref{figRik}.

\begin{figure*}[b]
\centering
  \includegraphics[width=0.5\textwidth]{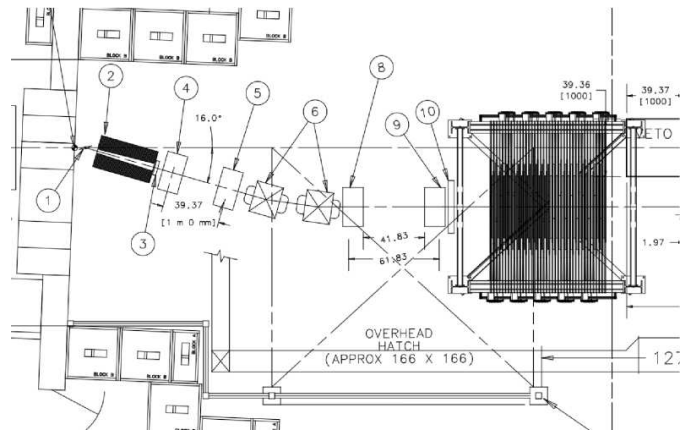}
  \caption{Drawing of the beamline at Meson Test Hall that shows all the elements used to characterize the tertiary beam. See above for an explanation of the elements.} \label{figDrawing}
\end{figure*}

\section{BeamLine Design}
The source of the beam is 120GeV protons coming from the Main Injector. This beam hits a 30 cm aluminium target and produces secondaries that will be bent through a magnet where we can select a particular secondary momentum. Most of the content of the beam is pions, but, as the selection is done using only magnets, we expect high electron content as well at low energies (up to 60\% for 4 GeV/c). Due to geometrical constraints we also expect a 1-5\%\cite{colemanref} of standard deviation in momentum distribution from the mean. \\We have chosen an 8 GeV/c beam; it will enter the MTest hall and interact with a secondary target. We designed the rest of the beamline so that it will transport particles of desired momentum of $300\,\mathrm{MeV/c}\,<\,p\,<\,1500\,\mathrm{MeV/c}$. The elements of this beamline are the following:
\begin{itemize}
\item[1.] A copper target, used to generate tertiaries from the 8GeV/c secondary beam;
\item[2.] A steel collimator, for tertiaries which also serves as a dump for the incoming beam;
\item[3,10.] Two time of flight (TOF) scintillator planes (120ps), for timing.
\item[4,5,8,9.] Four wire chambers (0.5mm), for angle measurements and tracking;
\item[6.] Two dipole magnets (10 IV 18), used as a spectrometer;
\end{itemize}

\begin{figure*}[t]
  \centering
  \includegraphics[width=0.5\textwidth]{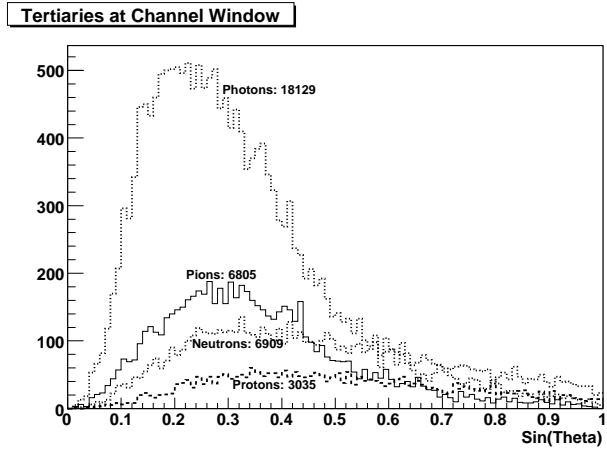}
  \caption{Angular distribution of tertiaries coming from the target and entering the channel at the collimator. For charged particles in this plot we have select momentum in the range $300\mathrm{MeV/c}\,<\,p\,<\,1500\mathrm{MeV/c}$.} \label{figTheta}
\end{figure*}

\begin{figure*}[b]
\centering
  \includegraphics[width=0.4\textwidth]{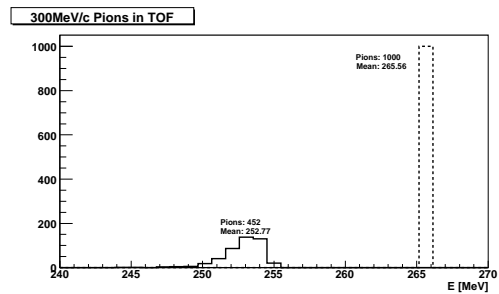}
  \caption{Scintillators, 20 cm x 20 cm x 2 cm separated by 6 m, are exposed to a 300 MeV/c pion beam. Dashed line: pions hitting the first face of the first scintillator counter; solid line: pions hitting the second face of the second scintillator. Besides the loss of almost 13 MeV in energy, note that there is about 55\% of the signal lost mostly due to multiple scattering in the plastic. In vacuum, we would reduce the loss by 5\%.} \label{figTOFs}
\end{figure*}

\begin{figure*}[t]
\centering
  \includegraphics[width=0.4\textwidth]{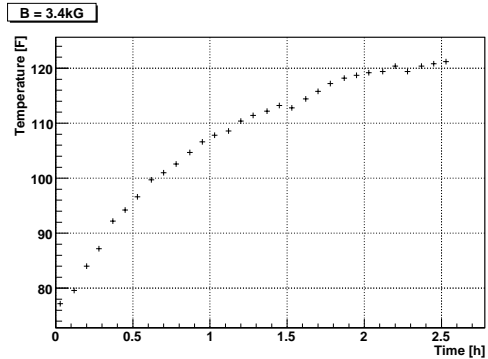}
  \caption{Temperature of magnet. The magnet is working below the limit temperature: 146 F.} \label{figDoug}
\end{figure*}

The tertiaries will trigger the first TOF detector and will traverse two wire chambers for angle measurement before entering the momentum analyzer magnets. Once bent, the beam will hit the last two wire chambers, the second TOF detector and finally the TBD; see figure \ref{figDrawing}.\\ The actual bending plus the time of flight will allow us to characterize the species that hits the detector and its momentum. The design of the geometry and the magnetic field however is far from trivial due to the spatial constraints of the hall.
\\ According to the Monte Carlo simulation\cite{g4beamlineref}, the production of low energy pions and protons is essentially flat from $10^\circ$ to $20^\circ$. $16^\circ$  is a good choice for our needs. Figure \ref{figTheta} shows the angular distribution rate of particles entering the collimator. \\ In order to maximize the yield, the shape of the channel within the collimator is chosen so that it matches the aperture of the magnets and the projection of the target. The position of the rest of the elements was also chosen to maximize the yield of the particles we are interested in. \\ According to figure \ref{figDrawing}, the beam will go through 4 cm of scintillator before reaching the main detector. That will reduce the yield of our low energy beam, up to 50\% for 300 MeV/c pions, due to multiple scattering as is shown in figure \ref{figTOFs}. \\ The temperature of the magnets have been tested for a 3.4kG central field; see figure \ref{figDoug}. The magnets behave correctly for this field.

\section{Analysis of the beam}
\begin{figure*}[t]
\centering
  \includegraphics[width=0.8\textwidth]{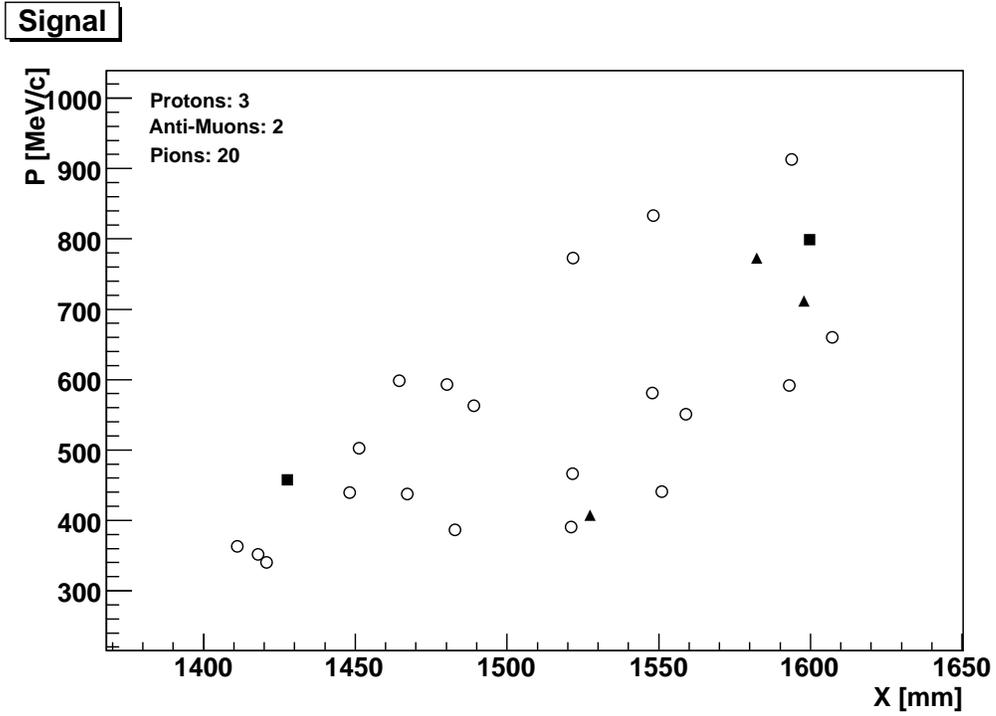}
  \caption{Momentum vs position of particles at the detector that have been seen by both TOF detectors (20 cm x 20 cm x 2 cm scintillator planes). Circles: $\pi^+$, squares: $\bar\mu$, triangles: proton. X measured from secondary beam axis.} \label{figTrigger}
\end{figure*}

\begin{figure*}[b]
\centering
  \includegraphics[width=0.8\textwidth]{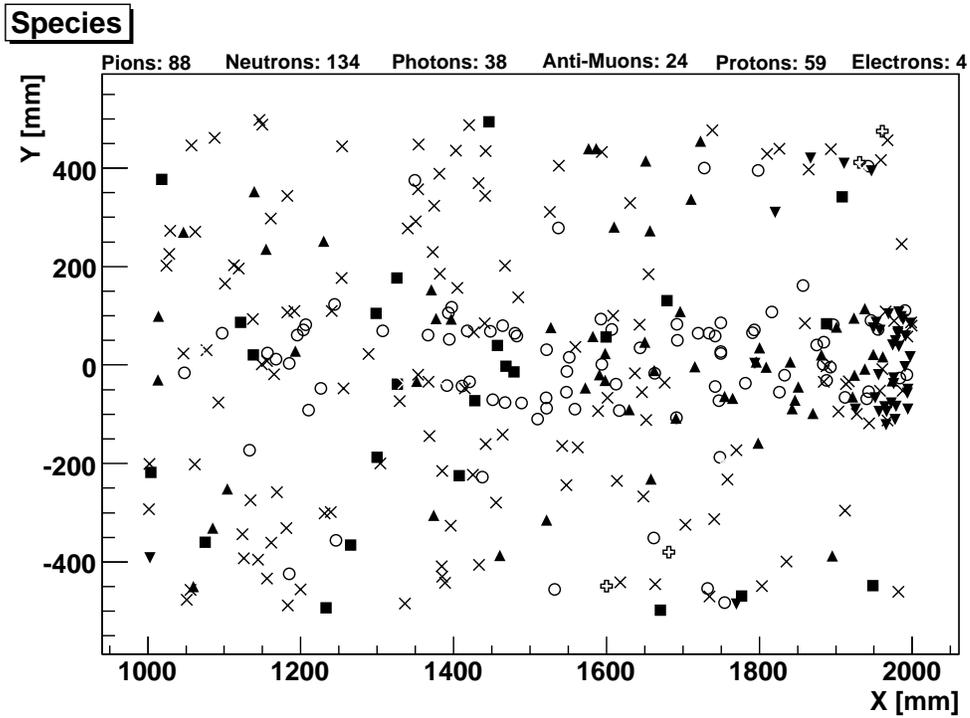}
  \caption{Composition of particles hitting the full Test Beam Detector after the cut in time. Circles: $\pi^+$, squares: $\bar\mu$, triangles: protons, reversed-triangles: photons, cross: electrons, equis: neutrons.} \label{figSpecies}
\end{figure*}

For this analysis we have taken out the TOF and wire chambers. We are interested only in the ratio of protons, $\mu^+$ and $\pi^+$ in the range of $300\,\mathrm{MeV/c}\,<\,p\,<\,1500\,\mathrm{MeV/c}$ momentum, called from now on {\it signal}, to the rest of beam content that hits the detector. Most of the electromagnetic background, and the higher energy punch-through in the collimator, will arrive faster than the actual {\it signal}, so we can make a cut in time using the time of flight system as a level zero trigger to get rid of this background. For a hundred thousand 8 GeV/c $\pi^+$ hitting the target we have 25 particles as {\it signal}, seen by TOF detectors, reaching the TBD. We have used a field of 3.5 kG for both magnets and the effects of fringe fields\cite{DanGreenref}. Figure \ref{figTrigger} shows the P-X correlation between these particles at the detector; besides the position there is also a correlation with angle that, by selecting incoming tracks, we can use to identify the {\it signal}. \\ Figure \ref{figSpecies} shows the particles that hit the full detector after the cut in time when no other dumping than the collimator is included. As most of the neutron content is produced just in the collimator, we considered appropriate to place a 10 cm concrete wall -paraffin or polyethylene should also work- just downstream the collimator to stop this low energetic undesirable contamination. In order to get rid of the remanent photon and electron content, we could use a thin layer of lead covering all the detector but the window of TOF2; this would not be thicker that a few radiation lengths.\\ We have used a spill of hundred thousand pions for the present Monte Carlo simulation, however the intensity of incoming secondary can be raised by a factor of up to 7.

\section{Conclusion}
Besides the present design is conceptually simple, it allows us to have complete control of the dispersion of the beam in order to fine-tune its energy. The combination of tracking chambers and time of flight detectors permit us to measure the position and the bending well enough to resolve 1 GeV/c particles with an error below $0.5\%$. In order to study the behaviour of the different configurations\cite{minervaref} of the TBD, this design allows us to trigger slow protons or fast pions and even to study electrons by changing the polarity in the magnets.

\section{Acknowledgments}
We want to thank David Boehnlein for all those useful discussions.

\end{document}